\documentclass[preprint,showpacs,preprintnumbers,amsmath,amssymb,nofootinbib]{revtex4}

\usepackage{graphics,amssymb,amsmath,amsthm,amscd,array}
\usepackage{graphicx,subfigure}

\usepackage{dcolumn}
\usepackage{bm}
\usepackage{amsmath,dsfont}
\usepackage{color}

\newcommand{\const}{\mathop{\rm const}\nolimits}

\def\p{{\partial}}

\def\vp{{\vec{p}}}

\def\vn{{\hat{n}}}

\def\vJ{{\vec{J}}}
\def\vL{{\vec{L}}}

\def\vK{{\vec{K}}}

\def\vC{{\vec{C}}}

\def\vN{{\vec{N}}}

\def\ort{{\rm o}}
\def\smallover#1/#2{\hbox{$\textstyle\frac{#1}{#2}$}}

\def\vK{{\vec{K}}}
\def\vJ{{\vec{J}}}

\def\hp{{\hat{p}}}
\def\vp{{\vec{p}}}
\def\vx{{\vec{x}}}

\def\vn{{\vec{n}}}

\def\vnabla{{\vec{\nabla}}}

\def\beq{\begin{equation}}
\def\eeq{\end{equation}}
\def\beqa{\begin{eqnarray}}
\def\eeqa{\end{eqnarray}}
\def\barray{\begin{array}}
\def\earray{\end{array}}
\def\nn{\nonumber}

\def\vTheta{{\vec{\Theta}}}

\def\vnabla{{\vec\nabla}}

\begin{document}

\preprint{arxiv:1006.1861}

\title{Non-commutative oscillator with Kepler-type  dynamical symmetry}

\author{\large P.~M.~Zhang \textsuperscript{a,b}}
\author{\large P.~A.~Horv\'athy \textsuperscript{a,b}}
\author{\large J.-P.~Ngome \textsuperscript{b}}
\affiliation{\textsuperscript{a} Institute of Modern Physics, Chinese Academy of Sciences, Lanzhou, China}
\affiliation{\textsuperscript{b}
Laboratoire de Math\'ematiques et de Physique Th\'eorique, 
Universit\'e de Tours,
F\'ed\'eration Denis Poisson - CNRS
Parc de Grandmont, 37200 Tours, France
}

\email{zhpm@impcas.ac.cn,juste.ngome-at-lmpt.univ-tours.fr, horvathy-at-lmpt.univ-tours.fr}
\bigskip
\bigskip
\begin{abstract}
 A $3$-dimensional non-commutative oscillator with 
no mass term but with a certain momentum-dependent potential admits a conserved Runge-Lenz vector,
derived from the dual description in momentum space.
The latter corresponds to a Dirac monopole with a fine-tuned inverse-square plus Newtonian potential, introduced by McIntosh, Cisneros, and by Zwanziger some time ago. The trajectories are (arcs of) ellipses, which, in the commutative limit, reduce to the circular
hodographs of the Kepler problem. The dynamical symmetry allows for an algebraic determination of the bound-state spectrum and actually extends to the conformal algebra o$(4,2)$.
\bigskip\\\bigskip
\noindent
Physics Letters {\bf A 374} (2010) 4275Ð4278
\end{abstract}


\maketitle


The ``Berry phase'' induced in the semiclassical analysis of a Bloch electron  in a 3-dimensional crystal 
makes the coordinates non-commutative  \cite{Niu},
\beqa
\{x_i,x_j\}=\epsilon_{ijk}\Theta_k,
\qquad
\{x_i,p_j\}=\delta_{ij},
\qquad 
\{p_i,p_j\}=0,
\eeqa
 where  $\Theta_i=\Theta_i(\vp)$.
Then the Jacobi identity requires the consistency condition $\vnabla_{\vp}\cdot\vTheta=0$~\footnote{For a more general theory which also includes magnetic fields, see, e.g.,
\cite{Niu,NCRev}. For simplicity, the mass has been chosen unity.}.
Choosing, for example, the non-commutative vector aligned in the third direction,
$
\Theta_i=\theta\delta_{i3},
\,
\theta=\const.
$,
 the $3D$ theory reduces to the planar mechanics
based on ``exotic'' Galilean sym\-metry
\cite{NCRev}. 
For the Kepler potential $V\propto r^{-1}$, for example, 
interesting results, including  perihelion precession, can be derived \cite{RoVe}. Other applications concern the
Quantum Hall Effect \cite{NCRev,DJTCS}.

Such a choice only allows for axial symmetry, though.  Full rotational symmetry can, however, be
restored by choosing instead $\vTheta$ to be a ``monopole in $\vp$-space'' \cite{BeMo},
\beq
\Theta_i=\theta\,\frac{p_i}{p^3}\,,
\qquad
\theta=\const.\,,
\label{monoptheta}
\eeq
where $p=|\vp|$\,. We record for further use that the associated symplectic structure has an extra, ``monopole'' term, $
\Omega=dp_i\wedge dx_i+(\theta/2p^3)\epsilon_{ijk}{p_i}dp_j\wedge dp_k,
$ 
that of a mass-zero, spin-$\theta$ orbit of the Poincar\'e [and indeed of the  $\ort(4,2)$  conformal] group \cite{CFH}.

In this Letter we study the $3D$ mechanics
with non-commutativity  (\ref{monoptheta}), augmented with a Hamiltonian of the form
$
H=
{p^2}/{2}+V(\vx,\vp).
$ 
Note that the potential may also depend on 
the momentum, $\vp$.
 The equations of motion read
\beq
\dot{x}_i={p_i}{}+\frac{\p V}{\p p_i}+\theta\epsilon_{ijk}\frac{p_k}{p^3}\frac{\p V}{\p x_j},
\qquad
\dot{p}_i=-\frac{\p V}{\p x_i}\, .
\eeq
Note, in the first relation, the  ``anomalous velocity terms''  due to our assumptions. 

We are particularly  interested in finding conserved quantities. This task is conveniently achieved by
 using
van Holten's covariant framework \cite{vH},
which amounts to searching for an expansion in
integer powers of the momentum,
$
Q=C_0(\vx)+C_i(\vx)p_i+ \frac{1}{2!}C_{ij}(\vx)p_ip_j+\frac{1}{3!}C_{ijk}(\vx)p_ip_jp_k+\dots
$ 
Requiring $Q$ to Poisson-commute with the Hamiltonian yields an infinite series of constraints.  The expansion can be truncated at a finite order $n$, however, provided the Killing equation is satisfied,
$ 
D_{(i_1}C_{i_2\dots i_n)}=0,
$  when we can set $C_{i_1\dots i_{n+1}\dots}=0$.

Let us assume that the potential has the form $V(r,p)$,
and try to find the conserved angular momentum, associated with the Killing vector
$ 
\vC=\vn\times\vx,
$ 
 which represents a space rotation around $\vn$.
An easy calculation shows, however,  that
the procedure fails to work, owing to the
``monopole'' term, which is not an integer power of $\vp$. 
We
propose therefore  to work instead  in ``dual'' (i.e. momentum) space,
and search for conserved quantities expanded 
rather into powers of the position,
\beq
Q=C_0(\vp)+C_i(\vp)x_i+\frac{1}{2!}C_{ij}(\vp)x_ix_j+
\frac{1}{3!}C_{ijk}(\vp)x_ix_jx_k\dots
\label{pConsQuant}
\eeq
Then van Holten's algorithm is replaced by
\begin{equation}
\begin{array}{llll}
C_i\left(\displaystyle{p_i}{}+ \displaystyle\frac{\p V}{\p p_i}\right)&=&0&\hbox{o(0)}
\\[12pt]
\displaystyle\frac{1}{r}
\displaystyle\frac{\p V}{\p r}\left(
\theta\epsilon_{ijk} \displaystyle\frac{p_k}{p^3}C_i
-\displaystyle\frac{\p C}{\p p_j}\right)
+C_{ij}\left(\displaystyle{p_i}{}
+\displaystyle\frac{\p V}{\p p_i}\right)&=&0
&\hbox{o(1)}
\\[12pt]
\displaystyle\frac{1}{r}
\displaystyle\frac{\p V}{\p r}\left(
\theta\displaystyle\frac{p_m}{p^3}\big(\epsilon_{ijm}C_{ik}
+\epsilon_{ikm}C_{ij}\big)
-\big(\displaystyle\frac{\p C_k}{\p p_j}+
\displaystyle\frac{\p C_j}{\p p_k}\big)
\right)
+C_{ijk}
\left(\displaystyle{p_i}{}
+\displaystyle\frac{\p V}{\p p_i}\right)&=&0\,
&\hbox{o(2)}
\\[12pt]
\displaystyle\frac{1}{r}
\displaystyle\frac{\partial V}{\partial r}\left(
\theta\displaystyle\frac{p_m}{p^3}
\big(\epsilon_{lim}C_{ljk}+\epsilon_{ljm}C_{lki}
+\epsilon_{lim}C_{lij}\big)
-\big(
\displaystyle\frac{\partial C_{ij}}{\partial p_k}+ \displaystyle\frac{\partial C_{jk}}{\partial p_i}+%
\displaystyle\frac{\partial C_{ki}}{\partial p_j}\big)\right)+
\nn
\\
\hskip50mm
C_{lijk}\left(\displaystyle{p_l}{}+
\displaystyle\frac{
\partial V}{\partial p_l}\right)&=&0
&\hbox{o(3)}
\\[12pt]
\vdots\qquad\qquad\qquad\qquad\qquad\vdots&\vdots
\end{array}
\label{constraints}
\end{equation}
where $r=|\vx|$. 
Then, for the dual Killing vector $\vC=\vn\times\vp
$, the algorithm provides us with the conserved angular
momentum,
\beq
\vJ=\vL-\theta\,\hp=\vx\times\vp-\theta\,\hp
\qquad
\hp={\vp}/{p},
\label{pAngMom}
\eeq
 which is what one would expect, due to the ``monopole in p-space", whereas the
NC parameter, $\theta$, behaves as the ``monopole charge'' \cite{Cortes}.

The next step is to inquire about second order conserved quantities. The usual Runge-Lenz vector is generated by
 the Killing tensor
$
C_{ij}=2\delta_{ij}\vn\cdot\vx-n_ix_j-n_jx_i,
$
where $\vn$ is some fixed unit vector \cite{vH}. 
Not surprisingly, the original procedure fails once again. The  dual procedure
(\ref{pConsQuant}) works, though. The two-tensor
\beq
C_{ij}=2\delta_{ij}\vn\cdot\vp-n_ip_j-n_jp_i,
\label{pRLK}
\eeq
 verifies the dual Killing equation of order 3. Then the order-2 equation yields
$
C_k=\theta\frac{(\vn\times\vp)_k}{p}.
$
Inserting into the first-order constraint and assuming
  $\p_rV\neq0$, the constraint is satisfied with $C=\alpha\vn\cdot\hp$,
\emph{provided} the potential and the Hamiltonian are
\beq
V=
\frac{\vx{}^2}{2}-\frac{p^2}{2}+\frac{\theta^2}{2p^2}
+\frac{\alpha}{p}\, ,
\qquad
H=\frac{\vx^2}{2}+\frac{\theta^2}{2p^2}+
\frac{\alpha}{p}\, ,
\label{Vee}
\eeq
respectively,
 where $\alpha$ is an arbitrary constant.
Then
the dual algorithm provides us with the 
Runge-Lenz-type vector
\beq
\vK=\vx\times\vJ-\alpha\hp
\,. 
\label{pRL}
\eeq
Its conservation can also be checked by a direct 
calculation, using the equations of motion,
\beq
\dot{\vx}=-\left(\frac{\theta^2}{p^4}+\frac{\alpha}{p^3}\right)\vp\
+\theta\,\frac{\vp\times{\vx}}{p^3},
\qquad
\dot{\vp}=-{\vx}{}.
\label{expleqmot}
\eeq

 Note that the $-p^2/2$ term in the potential cancels the usual kinetic term, and our system  describes a \emph{non-relativistic,
non-commutative particle with no mass term in an oscillator field, plus some momentum-dependent interaction}.
Writing the Hamiltonian as $H=\vx{}^2/2+(\theta^2/2)\big(p^{-1}+
\alpha/\theta^2\big)^2-\alpha^2/2\theta^2$ shows, moreover, that $H\geq -\alpha^2/2\theta^2$ with equality only attained when $p=-\theta^2/\alpha$, which plainly requires $\alpha<0$.
 
It is easy to understand the reason why our modified algorithm did work~: calling 
$\vp$ ``position'' and $-\vx$ 
``momentum'', the system can also be interpreted as an ``ordinary''
(i.e. massive and commutative) 
\emph{non-relativistic charged particle in the  field of a Dirac monopole of strength 
$\theta$, augmented with an inverse-square plus
a Newtonian potential}. This is 
the well-known ``McIntosh-Cisneros -- Zwanziger'' 
(MICZ) system \cite{MICZ,CFH},
for which the fine-tuned inverse-square potential is known to cancel
the effect of the monopole, allowing for a
 Kepler-type dynamical symmetry \cite{MICZ,CFH}. The angular momentum, (\ref{pAngMom}), and the
 Runge-Lenz vector, (\ref{pRL}), are, in particular, that  of the MICZ problem \cite{MICZ} in ``dual'' [momentum] space.

The conserved quantities provide us with valuable information on the motion. Mimicking what is done in the MICZ case, we note that
$\vJ\cdot\hp=-\theta$  implies 
that the vector $\vp$ moves on a cone of opening angle $\arccos(-\theta/J)$.
On the other hand, for
$ 
\vN={\alpha}\vJ-{\theta}\vK
$ 
we have $\vN\cdot\vp=\theta(J^2-\theta^2)=\theta L^2$
a constant, so that the
$\vp$-motion is in the plane perpendicular to $\vN$. \emph{The trajectory in $p$-space belongs therefore to a conic section}. 

For the MICZ problem, this is the main result, but
 for us here our main interest lies in finding the real space trajectories, $\vx(t)$. 
 By (\ref{expleqmot}), this amounts to finding the [momentum-] \emph{``hodograph''}   of the MICZ problem.
 Curiously, while the hodograph of the Kepler problem is well-known
 [it is a circle or a circular arc], we could not find the corresponding result in the (vast) literature of MICZ.  Returning to our notations, we note that due to  $\vN\cdot\vx=0$,
$\vx(t)$ also belongs to the same oblique plane, whose normal is $\vN$.

The problem is conveniently studied in an adapted coordinate system. One proves indeed that
\beq
\Big\{\hat{\imath},\hat{\jmath},\hat{k}\Big\}=
\Big\{\frac{1}{|\epsilon L|}\vec{K}\times\vec{J}
,\, \frac{1}{|\lambda \epsilon|}(2\theta H\vec{J}+\alpha \vec{K}),\,\frac{1}{|\lambda L|}(\alpha\vec{J}-\theta\vec{K})\Big\}
\eeq
[where $\lambda^2=\alpha^2+2H\theta^2,\;\epsilon^2=\alpha^2+2HJ^2$ and $L^2=J^2-\theta^2$]
is an orthonormal basis. (In $\hat{k}$ we recognize, in particular, $\vN/N$.)

Firstly, projecting onto these axis, $p_z=\vp\cdot\hat{k}=\theta L/|\lambda|=\const$, while for 
$p_x=\vec{p}\cdot\hat{\imath}$ and $p_y=\vec{p}\cdot\hat{\jmath}$ we find
\begin{equation}
\frac{\left(p_y+\frac{|\epsilon|\alpha}{2|\lambda|H}\right)^2}{\lambda^2/4H^2}-\frac{p_x^2}{L^2/2H}=1,
\end{equation}
which is the equation of a hyperbola or of an ellipse, depending on the sign of $H$. (For vanishing $H$ one gets a parabola). 
This confirms what is known for the MICZ problem \cite{MICZ}.

Next, for 
\beq
X=\vec{x}\cdot\hat{\imath}=-\frac{2|L|}{|\epsilon|}(H-\frac \alpha{2p}),
\qquad
Y=\vec{x}\cdot\hat{\jmath}=
-\frac{|\lambda|}{|\epsilon|}\frac{\vec{x}\cdot\vec{p}}p
\eeq
an easy calculation yields 
\begin{equation}
(X+\frac{|\epsilon|L}{J^2})^2+\frac{\alpha ^2L^2}{\lambda^2J^2}Y^2=
\frac{L^2\alpha^2}{J^4} \,.
\end{equation}
[completed with $Z=\vec{x}\cdot \hat{k}=0$]
which is an ellipse, since $\lambda^2=\alpha^2+2H\theta^2\geq0$.
The center has been shifted to $-{|\epsilon|L}/{J^2}$ along the
axis $\hat{i}$.
The major axis is directed along $\hat{\jmath}$.

Note that, unlike as in $\vp$-space, the $\vx$-trajectories are
always bounded.
When the energy is negative, $H<0$
[which is only possible when the Newtonian 
potential is attractive, $\alpha<0$], the whole ellipse is described. When $H>0$, 
[which is the only possibility in the repulsive case $\alpha>0$]
it is only the arc between the tangents drawn from the origin which is obtained. 
When the non-commutativity is turned off, $\theta\to0$, the known circular hodographs
of the Kepler problem are recovered. As $\alpha\to0$, the trajectory becomes unbounded, and follows the $y$-axis.
\begin{figure}
\begin{center}
\includegraphics[scale=.47]{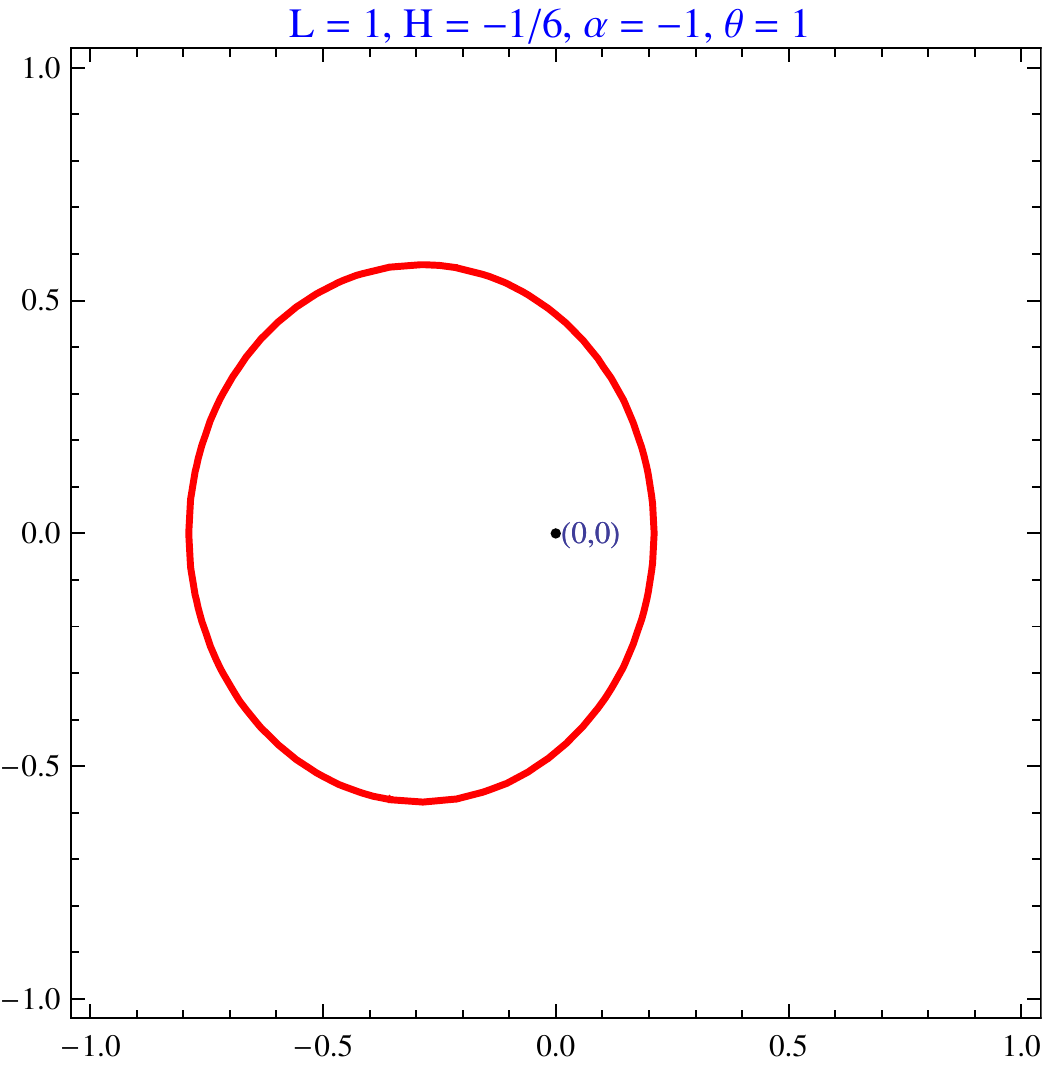}
\,
\includegraphics[scale=.47]{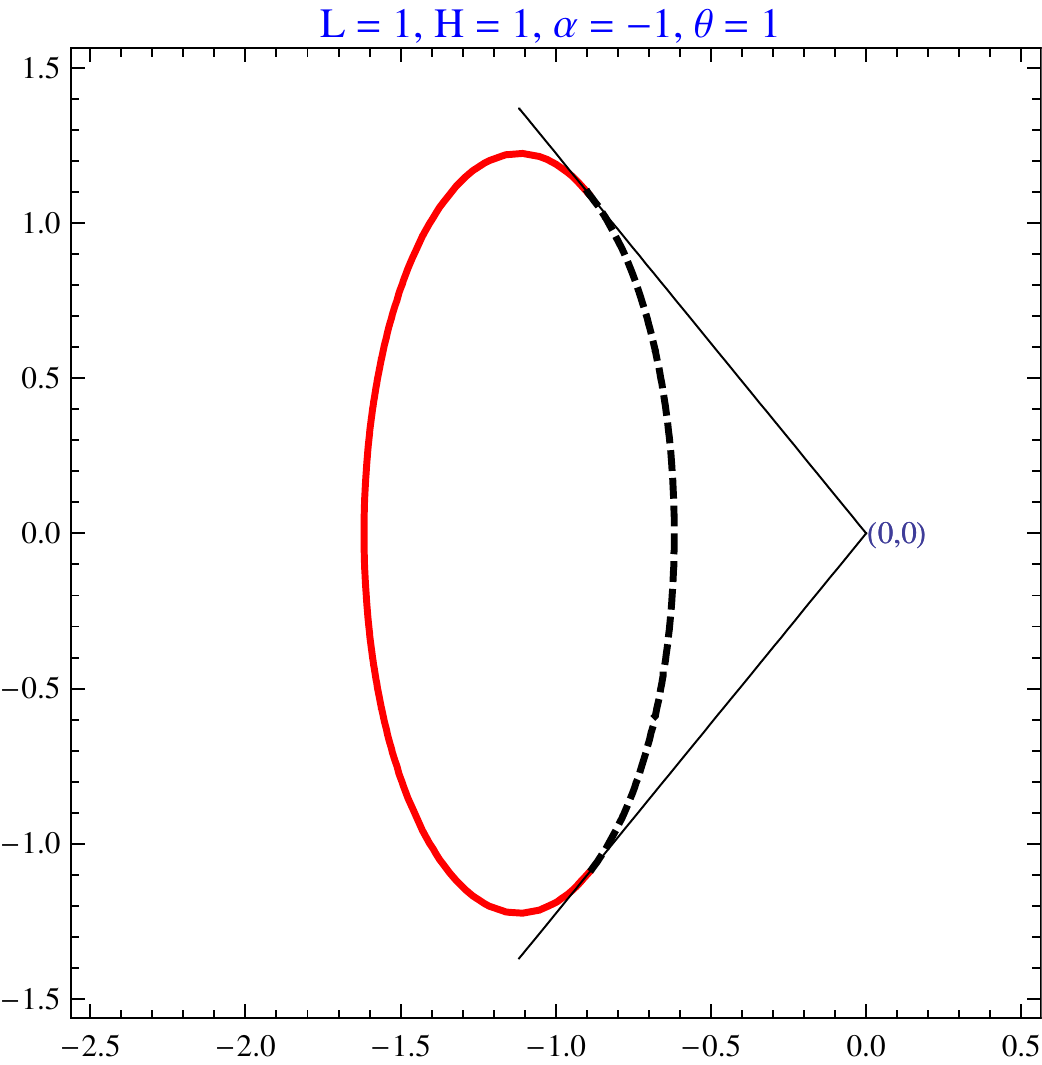}
\,
\includegraphics[scale=.47]{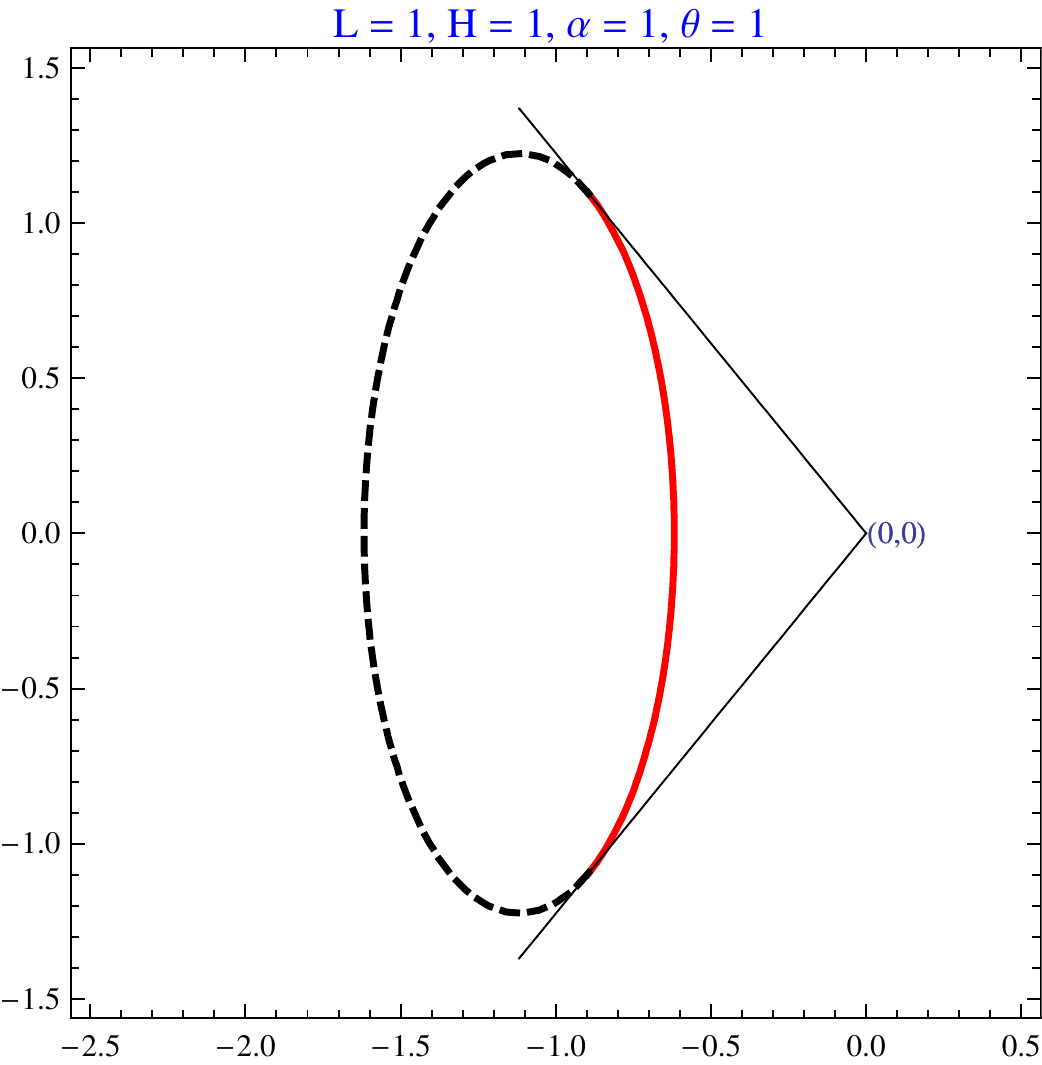}
\end{center}\vskip-5mm
\caption{\it When the Newtonian potential is attractive,
$\alpha<0$,
one can have $H<0$, and the trajectories are full ellipses.
The origin is inside the ellipse.
When $H>0$, which can arise either in the
 high-energy attractive or in the repulsive ($\alpha>0$) case, 
the origin is outside our ellipse.
The particle is confined to an arc [denoted with the heavy line] of the ellipse, determined by the tangents drawn from the origin. For $H=0$, the origin lies  on the ellipse, and 
``motion'' reduces to this single point. 
}
\label{PKplot}
\end{figure}

So far, we only discussed classical mechanics. Quantization is now straightforward using the known group theoretical properties of the 
MICZ problem in dual space. The non-commutativity [alias monopole charge] $\theta$ has to be an integer or  half
 integer -- this is indeed the first indication about the quantization of the NC parameter -- ;
wave functions should be chosen in the momentum representation, $\psi(\vp)$.
The angular momentum, $\vJ$, and the rescaled Runge-Lenz vector, $\vK/\sqrt{2|H|}$, close into 
$\ort(3,1)/\ort(4)$
depending on the sign of the energy.  In the last case, the representation theory provides us with the discrete energy spectrum [in units $\hbar=1$] 
\beq
E_n=-\frac{\alpha^2}{2n^2},
\qquad
n=n_r+\frac{1}{2}+(l+\frac{1}{2})\sqrt{1+\frac{4\theta^2}{(2l+1)^2}}\, ,
\eeq
where $n=0,1,\dots,\, l=0,1,\dots$, with degeneracy $n^2-\theta^2$.
The same result can plainly be derived directly by solving the Schr\"odinger equation in $\vp$-space \cite{MICZ}.

Moreover, the symmetry extends to the
conformal $\ort(4,2)$  symmetry, due to the fact that the massless Poincar\'e orbits with helicity $\theta$ are in fact orbits of the conformal group, cf. \cite{CFH}.


In conclusion, we observe that in most approaches one studies the properties (like trajectories, symmetries, etc.) of some given physical system. Here we followed in the reverse direction: after positing the fundamental commutation relations, we were 
looking for potentials with nice properties. This leads us 
 to  the  momentum-dependent potentials (\ref{Vee}), which is indeed a McIntosh-Cisneros --- Zwanziger system \cite{MICZ} in dual space. Unlike as in a constant electric field 
\cite{AHEHPA}, the motions lie in an (oblique) plane.
The particle is confined to bounded trajectories, namely 
to (arcs of) ellipses.

 The best way to figure our motions is to think of them as analogs of the circular hodographs of the Kepler problem to 
which they indeed reduce when the non-commutativity is turned off.
For $H<0$, for example, the dual motions are bound, and the velocity
turns around the whole ellipse; for $H>0$ instead,
our motion along a finite arc, starting from one extreme point
and tending to the other one at the end of the arc, corresponds to
the variation of the velocity in the course of a hyperbolic motion 
(of a comet, or in Rutherford scattering) in dual space.
   
There is also some analogy with particles with torsion \cite{Ptors}.

Our system, with monopole-type non-commutativity (\ref{monoptheta}), has some remarkable properties. 

Momentum-dependent potentials are rather unusual in high-energy physics. They are, however, widely used in nuclear physics, namely in the study of heavy ion collisions, where they correspond to non-local interactions  \cite{Nuphy}. Remarkably, in NC field theory, a $1/p^2$ contribution to the propagator emerges from UV-IR mixing. See the middle equation in equation (22) page 11 of Ref. \cite{Gubser}.

The absence of a mass term should not be thought of as the system being massless; it  is rather  reminiscent of ``Chern-Simons dynamics''  \cite{DJTCS}.

One can be puzzled how the system  would look like in configuration space. Trying to eliminate the momentum from the phase-space 
equations (\ref{expleqmot}) in the usual way, which amounts to deriving $\dot{\vx}$ w.r.t.
time and using the equations for $\dot{\vp}$,  fails, however,
owing to the presence of underived $\vp$  in the resulting equation.
This reflects  the non-local character of the
system.

One can, instead,  eliminate $\vx$ 
 using the same procedure, but in dual space.
This yields in fact the equations of motion of
MICZ in momentum space,
\begin{eqnarray}
\ddot{p}=\frac{J^2}{p^3}+\frac{\alpha}{p^2}\,,    
\qquad
\ddot{\vec{p}}=\frac{\alpha}{p^3}\,\vec{p}-\frac {\theta}{p^3}\,\vec{J}.
\label{ppeq}
\end{eqnarray}

Are these equations related to a theory with higher-order derivatives of the type \cite{LSZ}~?
The answer is yes and no. 
The clue is that \emph{time is not a ``good'' parameter}
for Kepler-type problems, owing to the impossibility of expressing it from the Kepler equation \cite{Cordani}. This is also the reason for which we describe the \emph{shape} of the trajectories, but we do not integrate the equations of motion. A ``better'' parameter can be found along
the lines indicated by Souriau
\cite{SouriauKepler,Bates} and then, deriving w.r.t. the new parameter, transforms
(\ref{ppeq}) into  a fourth-order linear matrix
differential equation, which can be solved.

It is, however, not clear at all if these
equations derive from some higher-order Lagrangian,
and if they happen to do, what would be the physical meaning of the latter.

The fourth-order equations do certainly \emph{not} come from one of the 
type in Ref. \cite{LSZ}~: the latter lives in fact in two space dimensions and has constant scalar non-commutativity $\theta$ --- while our system is 
3-dimensional and has a momentum-dependent vector $\vTheta(\vp)$, given in (\ref{monoptheta}).

It is tempting to ask if the relation to  the ``closest physical theory'' with a momentum - dependent potential, namely nuclear physics, can be
further developed and if similar (super)symmetries can be found also in nuclear physics. Once again, the answer seems to be negative, though~: while dynamical symmetries \emph{do}
play a role in nuclear physics \cite{Iachello},
those used so far do not seem to be of a momentum-dependent Keplerian types.

\begin{acknowledgments}\noindent
PAH is indebted to the {\it Institute of Modern Physics} of the Lanzhou branch of the {\it Chinese Academy of Sciences} for a special professorship, and PMZ is indebted to the {\it Laboratoire de Math\'ematiques et de Physique Th\'eorique}
of Tours University for hospitality.
JPN would like to thank the {\it R\'egion Centre} for a doctoral scholarship.  We are grateful to M. Plyushchay and to D. O'Connor for
useful correspondence. 
\end{acknowledgments}



\begin{thebibliography}{99}

\bibitem{Niu}
M. C. Chang and Q. Niu,
``Berry Phase, Hyperorbits, and the Hofstadter Spectrum.''
Phys. Rev. Lett. {\bf 75}, 1348 (1995).
For a review, see
  D.~Xiao, M.~C.~Chang and Q.~Niu,
``Berry Phase Effects on Electronic Properties,''
  arXiv:0907.2021 

\bibitem{NCRev}
It is impossible to give here an even approximately
complete list of relevant contributions.
The original references are 
C.~Duval and P. A. Horv\'athy,
``The ``Peierls substitution'' and the exotic Galilei group,''
 Phys. Lett. {\bf B 479}, 284-290 (2000).  [hep-th/0002233];
``Exotic galilean symmetry in the non-commutative plane, and the Hall effect,''  
Journ. Phys. {\bf A34}, 10097 (2001). [hep-th/0106089];
  V.~P.~Nair and A.~P.~Polychronakos,
  ``Quantum mechanics on the noncommutative plane and sphere,''
 Phys.\ Lett.\  B {\bf 505} (2001) 267  [arXiv:hep-th/0011172].
For a recent review, see, e.g.
  P.~A.~Horvathy, L.~Martina and P.~C.~Stichel,
  ``Exotic galilean symmetry and non-commutative mechanics,'' SIGMA {\bf 6} (2010), 060
  arXiv:1002.4772 
and references therein.

\bibitem{RoVe}
  J.~M.~Romero and J.~D.~Vergara,
  ``The Kepler problem and non commutativity,''
  Mod.\ Phys.\ Lett.\  {\bf A 18} (2003) 1673
  [arXiv:hep-th/0303064].
  
\bibitem{DJTCS}
  G.~V.~Dunne, R.~Jackiw and C.~A.~Trugenberger,
  ``Topological (Chern-Simons) Quantum Mechanics,''
  Phys.\ Rev.\  {\bf D 41} (1990) 661;
 In the non-commutative case, see P. A. Horv\'athy,
``The non-commutative Landau problem.''  
 Ann. Phys. (N. Y.) {\bf 299}, 128-140 (2002)
[hep-th/0201007].

\bibitem{BeMo}
  A.~B\'erard and H.~Mohrbach,
  ``Monopole in momentum space in noncommutative quantum mechanics,''
  Phys.\ Rev.\  {\bf D 69} (2004) 127701
  [arXiv:hep-th/0310167].
 The analogy with  
 anyon has been discussed in
 M. S. Plyushchay,
``Monopole Chern-Simons term: Charge monopole system as a particle with spin.''
Nucl. Phys. {\bf B589} 413 (2000), [hep-th/0004032] 

\bibitem{CFH}
P. A. Horv\'athy, B. Cordani and L. Feh\'er~:
``Kepler-type dynamical symmetries of long-range monopole interactions.'' 
 Journ. Math. Phys. {\bf 31}, 202 (1990).
 See also 
  M.~S.~Plyushchay,
  ``Massless point particle with rigidity,''
  Mod.\ Phys.\ Lett.\   {\bf A 4}, 837  (1989).

\bibitem{vH}
  J.~W.~van Holten,
  ``Covariant hamiltonian dynamics,''
  Phys.\ Rev.\  D {\bf 75}, 025027 (2007)
  [arXiv:hep-th/0612216].
  Using  Killing tensors has been advocated  by B. Carter, 
  ``Global structure of the Kerr family of gravitational fields,''
  Phys. Rev. {\bf 174}, 1559 (1968).
 For recent applications, see e.g., 
  P.~A.~Horvathy and J.~P.~Ngome,
  ``Conserved quantities in a non-abelian monopole field,''
  Phys.\ Rev.\  D {\bf 79} (2009) 127701
  [arXiv:0902.0273].
  J.~P.~Ngome,
  ``Curved manifolds with conserved Runge-Lenz vectors,''
  J.\ Math.\ Phys.\  {\bf 50} (2009) 122901
  [arXiv:0908.1204].
  M.~Visinescu,
  ``Higher order first integrals of motion in a gauge covariant Hamiltonian framework,''
  Mod.\ Phys.\ Lett.\  A {\bf 25} (2010) 341
  [arXiv:0910.3474].
  J.~P.~Ngome, P.~A.~Horvathy and J.~W.~van Holten,
  ``Dynamical supersymmetry of spin particle-magnetic field interaction,'' J. Phys. A: Math. Theor. {\bf A43} (2010) 285401,
  arXiv:1003.0137.
  T.~Igata, T.~Koike and H.~Ishihara,
  ``Constants of Motion for Constrained Hamiltonian Systems: A Particle around a Charged Rotating Black Hole,''
  arXiv:1005.1815 [gr-qc].
  
  
\bibitem{Cortes}
Jose L. Cortes, Mikhail S. Plyushchay,
``Anyons as spinning particles,''
Int.J.Mod.Phys. {A11} 3331 (1996). [hep-th/9505117]

  
\bibitem{MICZ}
H. V. McIntosh and A. Cisneros, 
``Degeneracy in the presence of a magnetic monopole.''
Journ. Math. Phys. (N.Y.) {\bf 11}, 896 (1970);
D. Zwanziger,
``Quantum field theory of particles with both electric and magnetic charges.''
Phys. Rev. {\bf 176}, 1489 (1968). 

\bibitem{AHEHPA}
P. A. Horv\'athy,
``The anomalous Hall effect in non-commutative mechanics,''
 Phys. Lett. {\bf A 359}, 705 (2006).
[cond-mat/0606472].

\bibitem{Ptors}
M. Plyushchay,
``Relativistic particle with torsion and charged particle in a constant electromagnetic field~: identity of evolution,''
Mod. Phys. Lett. {\bf A10}, 1463 (1995)
[hep-th/09309147].
   
\bibitem{Nuphy}
C. Gale, G. Bertsch, and S. Das Gupta,
``Heavy-ion collision theory with momentum-dependent 
interactions,''
Phys. Rev. {\bf C 35}, 1666 (1987);
C.B. Das, S.Das Gupta, Charles Gale, and Bao-An Li,
``Momentum dependence of symmetry potential in asymmetric
nuclear matter for transport model calculations,''
Phys. Rev. {\bf C 67} (2003) 034611;
``Effects of momentum-dependent symmetry potential on heavy-ion collisions induced by neutron-rich nuclei,''
Bao-An Li, Ch. B. Das, S. D. Gupta, Ch. Gale,
Nuclear Physics  {\bf  A 735} (2004)  563. 

\bibitem{Gubser}
S. S. Gubser and S. L. Sondhi,
``Phase structure of non-commutative scalar field theories,''
Nucl. Phys. {\bf  B605} 395 (2001) [hep-th/0006119].

\bibitem{LSZ}
~Lukierski J., ~Stichel P.~C. and  ~Zakrzewski W.~J.,
``Galilean-invariant $(2+1)$-dimensional models with a
    Chern-Simons-like term and $d=2$ noncommutative geometry,''
Annals of Physics  {\bf 260} (1997), 224,
[hep-th/9612017];\
 ``Noncommutative planar particle dynamics with gauge
interactions'', Ann.~Phys. (N.Y.) {\bf 306} (2003), 78 [hep-th/0207149].


\bibitem{Cordani}
B. Cordani,
{\it The Kepler Problem. Group Aspects, Regularization and Quantization. With an Application to the Study of Perturbations.} Birkhauser: Basel
(2003). 

\bibitem{SouriauKepler}
Jean-Marie Souriau,
``Global geometry of the two-body problem,'' (In French),
 CPT-82/P-1434, Jun 1982. 51pp. 
Presented at Symp. IUTAM-ISSM 'Modern Developments in Analytical Mechanics', Turin, Italy, Jun 1982.

\bibitem{Bates}
L. M. Bates,
``Geometric quantization of a perturbed Kepler problem,''
Rept. Math. Phys. {\bf 28} (1989)  289.

\bibitem{Iachello}
F. Iachello,
``Dynamic symmetries and supersymmetries in nuclear physics,''
Rev. Mod. Phys. {\bf 65}  (1993) 569.


\end{thebibliography}
\end{document}